\documentstyle[12pt]{article}
\title{Nuclear Field Theory and Chiral Symmetry on a Calabi-Yau Manifold}
\author{}
\date{}
\newcommand{\BEQ}{\begin{equation}}
\newcommand{\EEQ}{\end{equation}}
\newcommand{\BEQA}{\begin{eqnarray}}
\newcommand{\EEQA}{\end{eqnarray}}
\newcommand{\BA}{\begin{array}}
\newcommand{\EA}{\end{array}}
\newcommand{\BT}{\begin{tabular}}
\newcommand{\ET}{\end{tabular}}
\newcommand{\BC}{\begin{center}}
\newcommand{\EC}{\end{center}}
\newcommand{\DS}{\displaystyle}
\begin{document}
\maketitle
\noindent Author: J. Anthony de Wet \\
Comments: Contact jadew@pixie.co.za\\
Report no: NFT-97-03\\[5mm]
{\abstract The purpose of this contribution is to show how a nuclear field theory
follows naturally from the structure of four--dimensional Riemannian geometry.
A Yang--Mills field is introduced by constructing fibres that include all
possible exchanges of spin, parity and charge such that the collective
quantum numbers remain the same. In this way O(4) internal symmetry
transformations are found and a connection is obtained by exponentiation
of a CP--invariant operator C associated with the ground state. The metric is
Calabi--Yau and Einstein. \\
Carbon 13 is chosen as an example because it is the lightest nucleus to exhibit
small spin mutations even though there is no deformation parameter in the O(4)
commutation relations. Instead a supersymmetric transformation replaces a
quantum group. Mirror symmetry is also discussed.}
\section{Introduction}
de Wet (1996) considered an example of how a $Z_{2}$--graded algebra, specifically
the Lie algebra of $O(4)$, leads naturally to the well known angular momentum
matrices $\sigma_{i}$ of a coupled system of $P$ protons and N neutrons,
namely
\BEQ \sigma_{i}\:=\:E_{N}\:\otimes\: ^{P}\Gamma_{i}\:+\:^{N}\Gamma_{i}\:\otimes
\:E_{P},\:\:i=1,2,3\EEQ
where $\displaystyle{^{P}\Gamma_{i},^{N}\Gamma_{i}}$ are $(P+1)$-, $(N+1)$
-dimensional Lie operators of $so(3)$; $\displaystyle{E_{P},E_{N}}$ are $(P+1)$,
$(N+1)$ unit matrices. \\[5mm]
Essentially a $\displaystyle{Z_{2}}$-graded algebra splits a bundle $\DS{
\Lambda^{2}}$ into the direct sum
\BEQ\Lambda^{2}\:=\:\Lambda^{2}_{+}\:+\:\Lambda^{2}_{-}\EEQ
of self-dual and anti-self-dual $2$-forms respectively. An example of this
grading is the decomposition
\BEQ so(4)\:\cong \:so(3)\:+\:so(3) \EEQ
into bundles of three-dimensional Lie algebras which were long ago identified
by de Wet (1971) with spin and isospin (based upon some ideas of Eddington).
In a seminal paper Atiyah et al. (1978) uses this decomposition on the Lie
group level to introduce, at least locally, the two complex spinor bundles
$\DS{V_{+}}$ and $\DS{V_{-}}$ : the bundles of self-dual and anti-self-dual
spinors. Then $\DS{V\:=\:V_{+}\:+\:V_{-}}$ is isomorphic to the complexified
Clifford algebra bundles of one forms $\DS{\Lambda^{1}}$ (Eddington (1946)
called the Clifford algebra $\DS{C_{4}}$ a Sedenion algebra and we will
use his transparent Sedenion, or $E$-number, notation).\\[5mm]
The purpose of this contribution is to show how a nuclear field theory
follows naturally from the structure of four-dimensional Riemannian
geometry and to this end we shall consider the Hodge star mapping
\BEQ *\::\:\Lambda^{2}\:\rightarrow\:\Lambda^{2} \EEQ
as transforming a nucleus into its mirror image i.e. $(P,N)\rightarrow (N,P)$.

Under these conditions
\BEQ{\rm spin}(\sigma)\:\rightarrow\:{\rm spin}(\sigma)\:\::\:\:{\rm isospin}(T_{3})\:\rightarrow
\:-\:{\rm isospin}(T_{3})\EEQ
are the self--dual and the self--anti--dual forms. An example is given by the
first and fourth columns of Table I of section 3. (Here we have denoted
parity by p and the spin by s and in \S 2 we shall see how the nuclear
charge--spin--parity states are labelled by the partition $[\lambda_{1}\lambda_{2}
\lambda_{3}\lambda_{4}]$ of (A=N+P) and its four permutations that appear
in (18)).
Furthermore Atiyah et. al. (1978) consider the decomposition of the complexified
Clifford algebra bundle
$$\Lambda^{1}\:=\:\Lambda^{0}_{c}\:+\:\Lambda^{1}_{c}$$
such that the image of $\DS{V_{-}}$ in $\DS{\Lambda^{1}_{c}}$ is the
subspace $\DS{\Lambda^{1,0}}$ of $(1,0)$ forms that defines a complex
structure of the kind considered by de Wet (1995,96). Now a complex
manifold in turn decomposes into a sum of the spaces $\DS{\Lambda^{1,0}}$
and $\DS{\Lambda^{0,1}}$ of $(1,0)$ and $(0,1)$ forms (Kobayashi and
Nomizu(1969) ch. IX) so it is natural to identify the space $\DS{\Lambda^{0,1}}$
with the self--dual form $\sigma$ associated with $\DS{V_{+}}$. Then its conjugate
\BEQ\pi_{i}\:=\:E_{N}\:\otimes \:^{P}\Gamma_{i}\:-\:^{N}\Gamma_{i}\:\otimes\:
E_{P}\:\:\:(i=1,2,3)\EEQ
lies in $\DS{\Lambda^{1,0}}$ (Ibid). We shall see in \S 2 how $\pi$ is parity
but for the moment simply observe that this definition is also consistent
with Table I as described in \S 3. The six operators $\sigma_{i},\:\pi_{i}$ are
generators of $O(4)$. \\[5mm]
Now a complex structure occurs only on fermions (odd A), the even A nuclei being
characterised by shell structure, and an example of the decomposition of the
2 complex manifolds carrying $\DS{^{9}Li,\:^{9}C}$ is
\BEQ ^{9}Li\::\:6\:C_{[3303]}\:=\:34(\sigma+\pi)+9(\sigma\pi^{2}+\sigma^{2}\pi)+
(\sigma^{3}+\pi^{3})\EEQ
\BEQ ^{9}C\::\:6\:C_{[3033]}\:=\:34(\sigma-\pi)+9(\sigma\pi^{2}-\sigma^{2}\pi)+
(\sigma^{3}-\pi^{3})\EEQ
which is manifestly CP--symmetric because $\DS{T_{3}\rightarrow -T_{3}}$ is
accompanied by $\pi \rightarrow -\pi$. Equation (7) confirms the
decompositions given by Kobayashi and Nomizu (1969) and Salamon (1989) where
the Wigner coefficients are the number of times the irreducible spin
representations
\BEQ S^{1,1}\:=\:(\sigma +\pi),\:S^{2,1}\:=\:(\sigma\pi^{2}+\sigma^{2}\pi),\:
S^{3,0}\:=\:(\sigma^{3}+\pi^{3})\EEQ
are contained in the subspaces $\DS{\Lambda^{1,1}\:,\:\Lambda^{2,1}\:,\:
\Lambda^{3,0}}$ of $\DS{\Lambda^{3}}$ which are embedded in the Clifford
algebra of the A coordinates of $\sigma ,\:\pi$ and their products (cf.
Lawson and Michelsohn (1989) for the isomorphism between Clifford algebras
and exterior products). \\[5mm]
An inspection of (7), (8) shows clearly that fermion CP-invariance follows
from the decompositions $\DS{S^{1,1},\:S^{2,1},\:S^{3}}$ of the complex
manifold. Moreover since such a decomposition applies only to the state
$[3303]$ we will associate the ground state with the label $\DS{[\Lambda]
\equiv [\Lambda_{1}\Lambda_{2}\Lambda_{3}\Lambda_{4}]}$. Then higher energy
states will be labelled by $\DS{[\lambda]\equiv [\lambda_{1}\lambda_{2}\lambda_{3}
\lambda_{4}]}$. These, however, are characterized  by the decay of the
ground state so can no longer be in a complex manifold. \\[5mm]
In section 3 it will be shown that the fermion manifolds have a Ricci-flat
Kaehler metric and are therefore Calabi--Yau. Recently there have been
several studies of the mirror symmetry of Calabi--Yau spaces (cf. e.g.
Strominger et. al. (1996)) but the mirror nuclear manifolds appear to be
isomorphic. For example, in section 3 the matrix representations of the
CP--invariant operators $\DS{C_{[4324]},C_{[4234]}}$ of respectively
$\DS{^{13}C, :^{13}N}$ are identical up to interchange of rows and columns.
These representations are derived by substituting (1),(6) into the
equations (36),(37) (which are the analogues of (7),(8)) and their
rotational eigenvalues $\DS{C^{\prime}_{[\lambda]}}$ appear in the last
column of Table I. However we can substitute directly in (36),(37) using (38)
which is derived from a canonical labelling scheme suggested by (18) of
section 2. Again this labelling gives rise to an isomorphism with almost
identical rotational eigenvalues $\DS{C_{[\lambda]}}$ in the penultimate
column of Table I. \\[5mm]
In fact there are only tiny spin mutations (marked by asterisks) associated
with the states $[2533],\:[4333]]$ of $\DS{^{13}C}$. As discussed in section
3 these are believed to be due to Yang--Mills interaction even though the group
$O(4)$ has no deformation parameter $q$ in its commutation relations and is
not a quantum group. Instead interaction simply changes the spins of two
neutrons in paired states so we have replaced quantum group theory by
supersymmetry! \\[5mm]
In line with the aims of this contribution we have outlined several
correspondences between nuclear theory and the structure of $\DS{Z_{2}}$--
graded algebras which of course also plays a role in quantum group theory as
outlined by Manin(1991) chapter 4. We can now move on to how a Yang--Mills
field is incorporated.

\section{FIELD THEORY}
The basic theory has been reviewed in section 1 of de Wet (1994) so only an
outline will be given here. The method used constructs tensor products in the
enveloping algebra $A(\gamma)$ of the Dirac ring of an irreducible
self--representation
\BEQ\frac{1}{4}\Psi\:=\:(iE_{4}\psi_{1}+E_{23}\psi_{2}+E_{14}\psi_{3}+E_{05}
\psi_{4})e \EEQ
with itself. Here Eddington's E-numbers are related to to the Dirac matrices by
$$\gamma_{\nu}=iE_{\sigma\nu},\:E_{\mu\nu}=E_{\sigma\mu}E_{\sigma\nu},\:
E_{\mu\nu}^{2}=-1,\:E_{\mu\nu}=-E_{\nu\mu} \:\:\mu < \nu = 1,...,5$$
and the commuting operators $\DS{E_{23},\:E_{14},\:E_{05}}$ are respectively,
independent infinitesimal rotations in 3--space, 4--space and isospace that
correspond to the spin $\sigma$, parity $\pi$, and charge $\DS{T_{3}}$ carried
by a single nucleon. the parameters $\DS{\psi_{1},\:\psi_{2},\:\psi_{3}}$ are
half angles of rotation and $e$ is a primitive idempotent of the Dirac ring;
$\DS{E_{4}}$ is the unit matrix. \\[5mm]
A rotation of $180^{o}$ about $x$ will change spin up to spin down and if this
is followed by a rotation of $180^{o}$ about $t$, $x$ can go to $-x$ without
inverting time, but instead changing to a left--handed coordinate system system.
Thus we associate the rotation $\DS{E_{14}}$ about $x$ in 4--space with a parity
reversal $\DS{E_{14}\rightarrow -E_{14}}$, and this way the time coordinate
is `rolled up' so that the Lorentz-invariant representation (10) can
describe a nucleon in 3--space. \\[5mm]
The basis elements of $A(\gamma)$ are the $\DS{4^{A}\times 4^{A}}$ matrices
(A=N+Z)
$$E_{\mu\nu}^{l}\:=\:E_{4}\otimes\cdots\otimes E_{4}\otimes E_{\mu\nu}\otimes
E_{4}\otimes\cdots\otimes E_{4}$$
with $\DS{E_{\mu\nu}}$ in the $l$th position. The elements $\DS{E^{l}_{\mu\nu},\:
E^{l+1}_{\mu\nu}}$ commute, and $A(\gamma)$ is found to have the following
generators
\BEQ\Gamma^{(A)}_{\nu}\:=\:\frac{1}{2}(E^{1}_{0\nu}+E^{2}_{o\nu}+\cdots +
E^{A}_{0\nu}), \:\:\nu=1,\ldots ,5 \EEQ
\BEQ\sigma^{(A)}_{\mu\nu}\:=\:[\Gamma^{(A)}_{\mu},\Gamma^{(A)}_{\nu}]\:=\:(
E^{1}_{\mu\nu}+E^{2}_{\mu\nu}+\cdots +E^{A}_{\mu\nu})/2\EEQ
\BEQ\eta^{(A)}_{\nu}\:=\:E_{0\nu}\otimes\cdots E_{0\nu}\:=\:E^{1}_{0\nu}E^{2}_
{o\nu}\cdots E^{A}_{0\nu}\EEQ
\BEQ\eta^{(A)}_{\mu\nu}\:=\:\eta^{(A)}_{\mu}\eta^{(A)}_{\nu}\:=\:E^{1}_{\mu\nu}
E^{2}_{\mu\nu}\cdots E^{A}_{\mu\nu},\:\:\mu<\nu=1,\cdots,5.\EEQ
\\[5mm]
Then the irreducible representations, or minimal left ideals, of $A(\gamma)$
are
\BEQ\Psi^{(A)}\:=\:\sum_{\lambda}C_{[\lambda]} P_{[\lambda]}\EEQ
with
\BEQ C_{[\lambda]} = i^{\lambda_{1}}C(E^{1}_{23}\cdots E^{\lambda_{2}}_{23}
E^{\lambda_{2}+1}_{14}\cdots E^{\lambda_{2} + \lambda_{3}}_{14}E^{\lambda_{2}+
\lambda_{3}+1}_{05}\cdots E^{A-\lambda_{1}}_{05})\EEQ
if $C$ denotes summation over the $\DS{N_{[\lambda]}=A!/(\lambda_{1}!
\lambda_{2}!\lambda_{3}!\lambda_{4}!)}$ combinations of the basis elements
appearing in the bracket. Here $\DS{[\lambda]\equiv[\lambda_{1}\lambda_{2}
\lambda_{3}\lambda_{4}]}$ is a partition
\BEQ\lambda_{1}\:+\:\lambda_{2}\:+\:\lambda_{3}\:+\:\lambda_{4}\:=\:A
\EEQ
and
\BEQA P_{[\lambda]} & = & i^{-A}(i^{A}\psi^{\lambda_{1}}_{1}
\psi^{\lambda_{2}}_{2}\psi^{\lambda_{3}}_{3}\psi^{\lambda_{4}}_{4} + \eta
^{(A)}_{23}\psi^{\lambda_{1}}_{2}\psi^{\lambda_{2}}_{1}\psi^{\lambda_{3}}_{4}
\psi^{\lambda_{4}}_{3} \nonumber \\
& + & \eta^{(A)}_{14}\psi^{\lambda_{1}}_{3}\psi^{\lambda_{2}}_{4}
\psi^{\lambda_{3}}_{1}\psi^{\lambda_{4}}_{2} + \eta
^{(A)}_{5}\psi^{\lambda_{1}}_{4}\psi^{\lambda_{2}}_{3}\psi^{\lambda_{3}}_{2}
\psi^{\lambda_{4}}_{1})\epsilon_{A} \EEQA
is a projection operator satisfying
\BEQ P^{2}_{[\lambda]}\psi\:=\:P_{[\lambda]}\psi,\:\:\:\psi\equiv \psi_{1}\psi_{2}
\psi_{3}\psi_{4}.\EEQ
Also $\DS{\epsilon_{A}=e\otimes\cdots\otimes e=e^{1}e^{2}\cdots e^{A}}$ is a primitive
idempotent in $A(\gamma)$ so that (18) has the same form for $A$ nucleons
as the basic representation (10). \\[5mm]
By studying (18) it can be shown that a canonical labelling scheme associates
$\DS{(\lambda_{3}+\lambda_{4})}$ with the number of nucleons with a positive
charge (i.e. the first two terms represent a nucleus and the last two terms
its mirror image). $\DS{(\lambda_{2}+\lambda_{3})}$ the number with a given spin
$\sigma$, and $\DS{(\lambda_{2}+\lambda_{4})}$ with a particular parity $\pi$.
Thus each partition (17) represents a charge-spin-parity state of a nucleus, and
by choosing $4\times 4$ matrix representations for $\DS{E_{23},E_{14},E_{05}}$
and constructing fibres that include every possible exchange of spin, parity
and charge between nucleons such that the collective quantum numbers remain the
same, it may be shown that $\DS{C_{[\lambda]}}$ partitions beautifully into a
de Rham decomposition of isobaric multiplets. Under these conditions the
$\DS{4^{A}\times 4^{A}}$ matrices (11),(12) shrink to (1),(6) with rows
labelled by the fibres $[\lambda]$; where $\DS{\sigma_{1}=\sigma^{(A)}_{23},
\pi_{1}=\sigma^{(A)}_{14}}$ are two of the six generators $\DS{\sigma_{j},\pi
_{j}}$ of $O(4)$. \\[5mm]
In this way the nucleons interact by means of a Yang--Mills gauge field which
can be determined by calculating the connections in the fibre bundle. This
has been done by de Wet (1996) by exponentiating $\DS{C_{[\Lambda]}}$ and
finding the Ricci--flat Kaehler metric of the resulting Calabi--Yau space or
torus. In \S 3 this will be shown to be Einstein which ties in with the
ideas of Capovilla et. al. (1990) that say that an $SU(2)$ connection
characterises a solution of the source-free Einstein field equations. In
fact a compact $4$--manifold acted on by a group $SU(2)$ must be a Ricci--flat
torus (cf. Salamon (1989) p.106). In other words the nuclear metric is a
solution of the source-free Einstein equations! \\[5mm]
From another point of view we can regard the Dirac algebra as the infinitesimal
ring of Minkowski space and therefore as a tangent space to $4$--dimensional
space--time in the spirit of Ashtekar (1988). The representations of the tangent
space that give us the internal symmetries $\sigma,\pi,T$ are by construction a
soldering form (cf. Ashtekar et. al. (1988)) and exponentiation must necessarily
take us back to source--free Einstein space. \\[5mm]
Returning to (16) the bases of the form $\DS{\Lambda^{\lambda_{2},\lambda_{3}}}$
are contained in $\DS{C_{[\lambda]}}$ without the $\DS{E_{05}}$ elements which
as we shall see are needed only to characterise a particular member of an
isobaric multiplet. Thus in the next section, where an outline of the
decomposition (7),(8) is given, it will become clear that a new $(p,q)$ subspace
appears whenever the products $\DS{\sigma^{p}_{0}\pi^{q}_{0}}$ of
\BEQ\sigma_{0}=2\sigma_{1}=(E^{1}_{23}+\cdots +E^{A}_{23}),\:\:\:\pi_{0}=2\pi_{1}
=(E^{1}_{14}+\cdots +E^{A}_{14})\EEQ
contains terms with the same indices. Under these conditions $\DS{p+q\leq
\lambda_{2}+\lambda_{3}}$. \\[5mm]
Although a general nuclear state is labelled by $[\lambda]$, there is only one
state $\DS{[\Lambda]=[\lambda_{1}\lambda_{2}\lambda_{3}\lambda_{4}]}$ having
the decomposition (7) associated with the ground state. Then $[\Lambda]$
itself carries carries all the spin--parity states $[\lambda]$ of Table I.
These label the rows of a submatrix
\BEQ\mu\:=\:\left[ \BA{cc}   & B \\ -B & \EA \right] \EEQ
of $\DS{C_{[\Lambda]}}$ which has the holomorphic coordinates $\DS{z_{k}=
\underline{+}i\lambda_{k}}$ where $\DS{\lambda_{k}}$ is the eigenvalue
associated with $\DS{[\lambda]_{k}}$ by means of the correspondence (38) and
is real for the submatrix $B$. \\[5mm]
In fact $\DS{z_{k},\overline{z}_{k}}$ characterize the horizontal subspace
of a complex Grassmann or Kaehler manifold (Kobayashi and Nomizu (1969), Chapter
IX) and because it is also Ricci-flat and Kaehler it is a twistor space
(using the definition of Lawson and Michelsohn (1989) Chapter IV section 9). We
shall see in \S 3 how a metric is obtained.
\section{AN EXAMPLE: CARBON 13.}
In this section the ideas already outlined will be brought together with an
example that exhibits spin mutation and at the same time illustrates in more
detail how the decomposition (7),(8) of a complex manifold is obtained.

We begin by replacing (16) with
\BEQ C_{[\lambda]}=i^{\Lambda_{1}}\sigma_{0}^{\Lambda_{2}}\pi_{0}^{\Lambda_{3}}
T_{0}^{\Lambda_{4}}-\sum_{\lambda}i^{\lambda_{1}}\sigma_{0}^{\lambda_{2}}
\pi_{0}^{\lambda_{3}}T_{0}^{\lambda_{4}}\EEQ
where in addition to (20)
$$T_{0}\equiv 2\Gamma^{(A)}_{5}=(E^{1}_{05}+\cdots +E^{A}_{05}).$$
The real quantum numbers $s$, $p$ and $\DS{T_{3}=\frac{1}{2}(Z-N)}$ of spin,
parity and charge are
\BEQ\sigma_{0}\:=\:2is,\:\pi_{0}\:=\:2ip,\:\:T_{0}\:=\:2iT_{3}\:=\:i(Z-N)
\EEQ
which show how the quantum numbers of individual nucleons are additive. \\[5mm]
The summation contains all those terms arising from repeated indices \\
$\DS{E^{j}_{23}E^{j}_{23};\:E^{j}_{23}E^{j}_{14};\:E^{j}_{23}E^{j}_{05}}$;
$\DS{E^{j}_{14}E^{j}_{05}}$ that yield a single term according to the
multiplication table

\BEQ
\BT{c|ccc}  & $\DS{E^{j}_{23}}$ & $\DS{E^{j}_{14}}$ & $\DS{E^{j}_{05}}$ \\
\hline
$\DS{E^{j}_{23}}$ & $\DS{i^{2}}$ & $\DS{iE^{j}_{05}}$ & $\DS{iE^{j}_{14}}$ \\
$\DS{E^{j}_{14}}$ & $\DS{iE^{j}_{05}}$ & $\DS{i^{2}}$ & $\DS{iE^{j}_{23}}$ \\
$\DS{E^{j}_{05}}$ & $\DS{iE^{j}_{14}}$ & $\DS{iE^{j}_{23}}$ & $\DS{i^{2}}$
\ET \EEQ
\\
\noindent An elementary example is
\BEQ\sigma_{0}T_{0}\:=\:P(E^{i}_{23}E^{j}_{05})+i\pi_{0}\EEQ
where $P$ denotes summation over the $A!/(A-n)!$ permutations of the $n$
generators in the bracket. Then
\BEQ C_{[(A-2)101]}=i^{A-2}P(E^{i}_{23}E^{j}_{05})\:=\:i^{A-2}(\sigma_{0}T_{0}
-i\pi_{0}) \EEQ
and if $A=3,Z=1;\:T=-i$ so
\BEQ C_{[1101]}=(\sigma_{0}+\pi_{0})\EEQ
which characterizes the ground state of $^{3}H$. Now interchange $\DS{\sigma_{0}
\leftrightarrow \pi_{0}}$ in (26) to get
\BEQ C_{[(A-2)011]}=i^{A-2}(\pi_{0}T_{0}-i\sigma_{0})\EEQ
Then if $A=3,Z=2;\:T=i$ we have
\BEQ C_{[1011]}=(\sigma_{0}-\pi_{0})\EEQ
which characterizes the ground state of $^{3}He$. Equation (27) is the
irreducible spin representation $\DS{\Lambda^{1}}$ of (9) which occurs
once only and (25) is an example of a single term $\DS{\pi_{0}}$ arising
from $\DS{E^{j}_{14}=E^{i}_{23}E^{i}_{05}\:(i=1,\ldots,A)}$. Because
$\DS{T_{0}}$ is a scalar this term dictates the size of the subspace $\DS{
\Lambda^{1}}$. \\[5mm]
Let us now `add' another nucleon by multiplying (25) by $\DS{\pi_{0}=
(E^{1}_{14}+\cdots +E^{A}_{14})}$ to obtain
\BEQ\sigma_{0}\pi_{0}T_{0}\:=\:P(E^{i}_{23}E^{j}_{14}E^{k}_{05})+i\{P(E^{i}_{23}
E^{j}_{23})+P(E^{i}_{14}E^{j}_{14})+P(E^{i}_{05}E^{j}_{05})\}+Ai^{3}\EEQ
where
\BEQ\sigma^{2}_{0}=P(E^{i}_{23}E^{j}_{23})\:+\:Ai^{2};\:\:\pi^{2}_{0}=P(E^{i}
_{14}E^{j}_{14})\:+\:Ai^{2};\:\:T^{2}_{0}=P(E^{i}_{05}E^{j}_{05})\:+\:
Ai^{2} \EEQ
thus
\BEQ\BA{ccl} C_{[(A-3)111]} & = & i^{A-3}P(E^{i}_{23}E^{j}_{14}E^{k}_{05}) \\
 & = & i^{A-3}[\sigma_{0}\pi_{0}T_{0}-i(\sigma^{2}_{0}+\pi^{2}_{0}+T^{2}_{0}
 -3Ai^{2})-Ai^{3}]\EA \EEQ
 Then if $A=4,\:Z=2,\:T_{0}=0$
 \BEQ C_{[1111]}\:=\:\sigma^{2}_{0}+\pi^{2}_{0}+8 \EEQ
 which characterizes the ground state of $\DS{^{4}He}$ found to have only
 one spin configuration. In this case there is no mirror nucleus and $A$ is
 even so there is no decomposition like that of (7),(8). We are in fact in a
 vertical subspace $h$ of the tangent space to the boson manifold with
 a matrix representation
 \BEQ \left[ \BA{cc} A &  \\   & C \EA \right] \EEQ
Clearly the process may be continued until ultimately the invariant operator
for $\DS{^{9}Li}$ is
\BEQ C_{[3303]}=\:i^{3}P(E^{i}_{23}E^{j}_{23}E^{k}_{23}E^{l}_{05}E^{m}_{05}
E^{n}_{05})/(3!3!)\EEQ
which yields (7) after writing $T=i(Z-N)=-3i$ and making use of subsidiary
relations such as (31).

When $A=13$ we find
\BEQ\BA{ccl} ^{13}C:\:-12C_{[4324]} & = & 3089(\sigma_{0}-\pi_{0})+
151(\sigma_{0}\pi^{2}_{0}-\sigma^{2}_{0}\pi_{0}) \\
& + & 135(\sigma^{3}_{0}-\pi^{3}_{0}) + 3(\sigma^{3}_{0}\pi^{2}_{0}-
\sigma^{2}_{0}\pi^{3}_{0}) \\
 & + & (\sigma_{0} \pi^{4}_{0}-\sigma^{4}_{0}\pi_{0})+(\sigma^{5}_{0}-
 \pi^{5}_{0})\EA \EEQ
\BEQ\BA{ccl} ^{13}N:\:-12C_{[4234]} & = & 3089(\sigma_{0}+\pi_{0})+151(\sigma_{0}
\pi^{2}_{0}+\sigma^{2}_{0}\pi_{0}) \\
& + & 135(\sigma^{3}_{0}+\pi^{3}_{0}) + 3(\sigma^{2}_{0}\pi^{3}_{0} \\
& + & \sigma^{3}_{0}\pi^{2}_{0})+(\sigma^{4}_{0} \pi_{0}+
\sigma_{0}\pi^{4}_{0})+(\sigma^{5}_{0}+\pi^{5}_{0})\EA \EEQ
and once again we have precisely the decomposition given by Salamon (1989, p33)
of $\DS{\Lambda^{5}=\Lambda^{\Lambda^{2}+\Lambda^{3}}}$. \\[5mm]
Now if we assume, in accord with the canonical labelling suggested by (18),
that $\DS{(\lambda_{2}+\lambda_{3})}$ is the number of nucleons with negative
spin $\sigma$ and $\DS{(\lambda_{2}+\lambda_{4})}$ that number with positive
parity $\pi$ then we can determine the eigenvalues of $\DS{C_{[\lambda]}}$
associated with each configuration $[\lambda]$ simply by substitution of
\BEQ\sigma_{0}\:=\:\{A-2(\lambda_{2}+\lambda_{3})\}i,\:\:\pi_{0}\:=\:\{2
(\lambda_{2}+\lambda_{4})-A\}i \EEQ
in (36),(37). These are eigenvalues without any interaction because as yet no use
has been made of (1),(6). However we can also substitute directly from
these equations (remembering from (20) that $\DS{\sigma_{0}=2\sigma_{1},\:\pi
_{0}=2\pi_{1}}$) and use the standard representations of $so(3)$ for $\DS{\Gamma
^{i}}$ to find a matrix representation $\mu$ of $\DS{C^{\prime}_{[\lambda]}}$.
The matrix representations of $\DS{^{13}N}$ and $\DS{^{13}C}$ are identical
up to an exchange of rows and columns and Table I (which does not show repeated
eigenvalues) compares the eigenvalues of $\DS{C_{[\lambda]}}$ and $\DS{C^{\prime}
_{[\lambda]}}$. Because of the parity change columns 1 and 4 will also yield
the same eigenvalues, as will columns 2 and 3 up to a sign change caused by
$\DS{\sigma_{0}\rightarrow -\sigma_{0}}$. Those states associated with the
matrix representation $\DS{C^{\prime}_{[\lambda]}}$ in the last column are
marked by an asterisk and have repeated eigenvalues except when $\DS{\lambda_{3}
=\lambda_{4}=3}$. \\[5mm]
It is apparent that only the spin-parity states $[2333]$, $[4333]$ exhibit a
tiny mutation of 2/900. However if the eigenvalues of these states are interchanged so that the
ground state $[4333]$ has the value -460 instead of -468, and $[2533]$
assumes 468 not 460 the mutations disappear. Thus these two states are
paired, differing only in the number of neutrons with negative spin, so
the introduction of a Yang--Mills field simply changes the spin of the two
neutrons in the paired states which amounts to a supersymmetric transformation.
Another example of paired states is given by Fig. 1 of de Wet (1995,96) where
$\DS{X_{4}}$ is the the ground state $[3303]$ of $\DS{^{9}Li}$ and $\DS{X_{5}}$
could be the state $[3321]$. \\[5mm]
Yang--Mills fields do not change the energy so there can be no dissipation
due to spin--mutations (this would lead to the collapse of nuclei to a
zero--spin state). Thus there must either be supersymmetry or the mutation
is carried by nucleons moving on two--dimensional toroidal surfaces in such a
way as to be anyons. \\[5mm]
To find the Kaehler metric on the fermion manifolds we need first to find
$\DS{exp(C_{[\lambda]}\theta)}$ which has been treated in some detail by
de Wet (1994, 1995, 1996). Specifically
\BEQ e^{\mu\theta}\:=\:\mu\sum^{n}_{k=0,1,..}\frac{F_{k}(\mu)cos\:\lambda_{k}\theta}
{i\lambda_{k}F_{k}(i\lambda_{k})}\:+\:i\sum^{n}_{k=1,2,..}\frac{F_{k}(\mu)}
{F_{k}(i\lambda_{k})}\:sin\:\lambda_{k}\theta \EEQ
where $\mu$ is an irreducible subspace containing  $[\Lambda]$ of $\DS{C_{
[\Lambda]}}$, $\DS{\{1;\lambda_{2};\ldots,\lambda_{n}\}}$ are normalised
positive, distinct and real eigenvalues of the subspace $B$ of (21), and
$$F_{0}(\mu)\equiv F(\mu)/\mu,\:\:F_{k}(\mu)\equiv F(\mu)/(\mu^{2}+\lambda_{k}^{2}),
\:\:F_{k}(\mu)F_{l}(\mu)=0$$
if
\BEQ F(\mu)\equiv \mu(\mu^{2}+1)(\mu^{2}+\lambda^{2}_{2})\cdots (\mu^{2}+\lambda
_{n}^{2})=0\EEQ
Writing (39)
$$e^{\mu\theta}=Z_{0}(cos\:\theta)+Z_{1}(sin\:\theta)=\left[\BA{rr}
Z_{0} & Z_{1} \\ -Z_{1} & Z_{0} \EA \right]$$
we can now follow Kobayashi and Nomizu(1969) Chapter IX \S 6 and Wong(1967)
to find the metric on a complex Grassmann manifold, i.e.
\BEQ ds^{2}\:=\:Tr \frac{dT}{(1+T\overline{T}^{t})} . \frac{d\overline{T}^{t}}
{(1+T\overline{T}^{t})} \EEQ
Where
\BEQ T\:\equiv \:Z_{1}Z^{-1}_{0}\:=\:-T^{t}\:=\:\mu\:\sum_{k=1,2,..}^{n}
\frac{i(F_{k}(\mu)/\mu)}{F_{k}(i\lambda_{k})}\:tan \lambda_{k}\theta \EEQ
\BEQ T\overline{T}^{t}\:=\:\sum_{k=1,2,..}^{n}\:K_{k}(\mu)tan^{2}\lambda_{k}\theta
\EEQ
\\
Here $\DS{\overline{T}^{t},\:d\overline{T}^{t}}$ are the conjugate transposes
of $T,\:dT$ and
\BEQ K_{k}(\mu)\:=\:i\lambda_{k}F_{k}(\mu)/F_{k}(i\lambda_{k})\mu \EEQ
is idempotent, so that (41) reduces to the flat measure carried by a torus,
namely
\BEQ ds^{2}\:=\:\sum_{k=1,2,..}^{p}dz_{k}d\overline{z}_{k}\:,\:\:z_{k}=i\lambda_{k}
\theta \EEQ
\\
However a translation to the normalized canonical form
\BEQ\{0;1;\lambda_{2};\ldots;\lambda_{n}\}\:\:\:n\leq p \EEQ
where $\{\lambda_{2};\ldots;\lambda_{n}\}$ are all positive, involves adding
or subtracting an angular momentum $\DS{\lambda_{0}}$ and then dividing by
$\DS{\lambda_{f}=(\lambda_{1}\underline{+} \lambda_{0}})$ which may be absorbed
in $\theta$ and does not change the geodesics although there is a frequency
change in the wave function $\DS{e^{\mu\theta}}$. Examples of the translation
are the last columns of Table I. \\[5mm]
The effect of the translation is to multiply (42) by $\DS{tan\:\lambda_{0}
\theta}$ which introduces the new distorted metric
\BEQ\BA{ccl} ds^{2} & = & g_{k\overline{k}}d(\lambda_{k}\theta)d(-\lambda_{k}\theta) \\
  & = & d(\lambda_{k}\theta)d(-\lambda_{k}\theta)\sum_{k}\frac{\mu}{\lambda_{k}}
  K_{k}(\mu)g(\lambda_{k}\mu)\sum_{k}\frac{\overline{\mu}^{t}}{\lambda_{k}}
  K_{\overline{k}}(\overline{u}^{t})g(-\lambda_{k}\theta)\EA \EEQ
with
$$g(\lambda_{k}\theta)=-g(-\lambda_{k}\theta)=tan\:\lambda_{0}\theta\:sec^{2}
(\lambda_{k}\theta)/(1+tan^{2}\lambda_{0}\theta\:tan^{2}\lambda_{k}\theta).$$
Here $\DS{\mu=-\overline{\mu}^{t},\:\:K_{\overline{k}}(\overline{\mu}^{t})=
K_{k}(\mu)}$ and $\DS{k=\lambda_{k}\theta}$ are the $p$ distinct coordinates
of $B$ in (21), $\DS{\overline{k}=-\lambda_{k}\theta}$ and $\DS{\underline{+}i
\lambda_{k}}$ are the coordinates of $\mu$. \\[5mm]
The metric (47) was used by de Wet (1996) to find peaks or horns on the
manifold of $\DS{^{9}Li}$ which could represent instantons that become quarks
or leptons at energies sufficiently high to break Yang--Mills symmetry; but
for the purposes of this contribution we simply observe that (47) is
Einstein according to the definition of Atiyah et. al. (1978) because only
even products of $\mu$ occur which means a diagonal representation like
(34). In other words the fermion metric is a solution of the source--free
Einstein equations. \\[5mm]
This also ensures that the Ricci tensor vanishes but the sectional curvature
$$K\:=\;R_{k\overline{k} k\overline{k}}\:=\:\frac{\partial^{2}g_{k\overline{k}}}
{\partial k \partial \overline{k}}\:-\:\sum^{p}\frac{\partial^{2}g_{l\overline{l}}}
{\partial l \partial \overline{l}}$$
does not because curvature is determined by the orientation of the remaining
p--planes. Thus a spinor field corresponding to the state $\DS{[\lambda]_{k}}$
and propagated parallelly only around the section $k\overline{k}$ will return
to its original value which is precisely the condition found by Green et. al.
(1993, chapter 15) to show that a Calabi--Yau space or $\DS{K^{3}}$ surface
carries a string field.\\[5mm]
\noindent {\bf ACKNOWLEDGEMENT} \\
It is a pleasure to thank Peter Hall of the University of Port Elizabeth
for his help.

\newpage

\noindent {\bf REFERENCES}

\vspace{8mm}
\noindent Ashtekar, A. (1988). New Perspectives in Canonical Gravity, Bibliopolis,
\\[3mm]
\hspace*{20mm} Naples.\\[3mm]
\noindent Ashtekar, A., Jacobson, T. and Smolin, L. (1988). Communications in \\[3mm]
\hspace*{20mm} Mathematical Physics , \underline{115}, 631.\\[3mm]
\noindent Atiyah, M.F., Hitchin, N.J. and Singer, I.M. (1978). Proceedings of the
\\[3mm] \hspace*{20mm} Royal Society of London, \underline{362}, 425 \\[3mm]
\noindent Capovilla, R., Jacobson, T. and Dell, J. (1990). Classical and Quantum
\\[3mm] \hspace*{20mm} Gravity, \underline{7}, L1-L3.\\[3mm]
\noindent de Wet, J.A. (1971). Proceedings of the Cambridge Philosophical Society,
\\[3mm] \hspace*{20mm} \underline{70}, 485.\\[3mm]
\noindent de Wet, J.A. (1994). International Journal of Theoretical Physics,
\underline{33}, 1841. \\[3mm]
\noindent de Wet, J.A. (1995). International Journal of Theoretical Physics,
\underline{34}, 1065. \\[3mm]
\noindent de Wet, J.A. (1996). International Journal of Theoretical Physics,
\underline{35}, 1201. \\[3mm]
\noindent Eddington, A.S. (1948). Fundamental Theory, Cambridge University Press,
\\[3mm] \hspace*{20mm}Cambridge. \\[3mm]
\noindent Green, M.B., Swartz, J.H. and Witten, E. (4th Ed. 1993). Superstring
\\[3mm] \hspace*{20mm} Theory, Cambridge University Press, Cambridge. \\[3mm]
\noindent Lawson, H.B. and Michelsohn, M.--L. (1989). Spin Geometry, Princeton
\\[3mm] \hspace*{20mm} University Press, Princeton,New Jersey. \\[3mm]
\noindent Kobayashi, S. and Nomizu, K. (1969). Foundations of Differential
Geometry, \\[3mm] \hspace*{20mm} Wiley--Interscience, New York. \\[3mm]
\noindent Manin, Y.I. (1991). Topics in Non--Commutative Geometry, Princeton
\\[3mm] \hspace*{20mm} University Press, New Jersey. \\[3mm]
\noindent Salamon, S. (1989). Riemannian Geometry and Holonomy Groups,
\\[3mm] \hspace*{20mm} Longmans Scientific, Essex. \\[3mm]
\noindent Strominger, A., Yau, S.--T. and Zaslow, E. (1996). Nuclear Physics,
\\[3mm] \hspace*{20mm} \underline{B479}, 243.
\newpage
{\bf TABLE 1.}\hspace{8mm}Coherent States of $\DS{^{13}C,\:^{13}N}$.
\footnotesize{
\BC\BT{|ll|ll|rr|rr|c|c|} \hline \\
\multicolumn{2}{c}{$^{13}C$} & \multicolumn{2}{c}{$^{13}N$} & & & & & &
\tiny{Matrix} \\ \cline{1-4}
\multicolumn{2}{c}{s\hspace{4mm}+\hspace{4mm}-} & - & + & \multicolumn{2}
{c}{$\DS{^{13}C}$} & \multicolumn{2}{c}{$\DS{^{13}N}$} & &
\tiny{Representation} \\
\multicolumn{2}{c}{p\hspace{4mm}-\hspace{4mm}+} & - & + & \multicolumn{2}{c}
{\tiny{\underline{$\lambda_{1}\lambda_{2}\lambda_{3}\lambda_{4}$}}} &
\multicolumn{2}{c}{\tiny{\underline{$\lambda_{4}\lambda_{3}\lambda_{2}
\lambda_{1}$}}} & \underline{$\DS{C_{[\lambda]}+3500}$} & \underline{$\DS{C^{\prime}_{[\lambda]}
+3500}$} \\
\tiny{$\lambda_{1}\lambda_{2}\lambda_{3}\lambda_{4}$} & \tiny{$\lambda_{2}\lambda_{1}\lambda_{4}
\lambda_{3}$} & \tiny{$\lambda_{3}\lambda_{4}\lambda_{1}\lambda_{2}$} & \tiny{$\lambda_{4}
\lambda_{3}\lambda_{2}\lambda_{1}$} & $\sigma_{0}$ & $\pi_{0}$ &
$\sigma_{0}$ & $\pi_{0}$ & 3600 & 3600 \\ \hline
$7006$ & $0760^{*}$ & $0760^{*}$ & $6007$ & $13i$ & $-i$ & $13i$ & $i$ &
$35/9$ & $35/9$ \\
$7015^{*}$ & $0751$ & $1570$ & $5107^{*}$ & $11i$ & $-3i$ & $11i$ & $3i$ &
$0$ & $0$ \\
$7024$ & $0742^{*}$ & $2470^{*}$ & $4207$ & $9i$ & $-5i$ & $9i$ & $5i$ &
$1.4$ & $1.4$ \\
$7033^{*}$ & $0733$ & $3370$ & $3307^{*}$ & $7i$ & $-7i$ & $7i$ & $7i$ &
$56/90$ & $56/90$ \\ \hline
$6106^{*}$ & $1660$ & $0661$ & $6016^{*}$ & $11i$ & $i$ & $11i$ & $-i$ &
$5/9$ & $5/9$ \\
$6115$ & $1651^{*}$ & $1561^{*}$ & $5116$ & $9i$ & $-i$ & $9i$ & $i$ &
$2/3$ & $2/3$ \\
$6124^{*}$ & $1642$ & $2461$ & $4216^{*}$ & $7i$ & $-3i$ & $7i$ & $3i$ &
$114/90$ & $114/90$ \\
$6133$ & $1633^{*}$ & $3361^{*}$ & $3316$ & $5i$ & $-5i$ & $5i$ & $5i$ &
$68/90$ & $68/90$ \\ \hline
$5206$ & $2560^{*}$ & $0652^{*}$ & $6025$ & $9i$ & $3i$ & $9i$ & $-3i$ &
$1$ & $1$ \\
$5215^{*}$ & $2551$ & $1552$ & $5125^{*}$ & $7i$ & $i$ & $7i$ & $-i$ &
$10/9$ & $10/9$ \\
$5224$ & $2542^{*}$ & $2452^{*}$ & $4225$ & $5i$ & $-i$ & $5i$ & $i$ &
$1$ & $1$ \\
$5233$ & $2533^{*}$ & $3352^{*}$ & $3325$ & $3i$ & $-3i$ & $3i$ & $3i$ &
$99/90^{*}$ & $99.2/90$ \\ \hline
$4306^{*}$ & $3460$ & $0643$ & $6034^{*}$ & $7i$ & $5i$ & $7i$ & $-5i$ &
$86/90$ & $86/90$ \\
$4315$ & $3451^{*}$ & $1543^{*}$ & $5134$ & $5i$ & $3i$ & $5i$ & $-3i$ &
$88/90$ & $88/90$ \\
$4324^{*}$ & $3442$ & $2443$ & $4234^{*}$ & $3i$ & $i$ & $3i$ & $-i$ &
$79.6/90$ & $79.6/90$ \\
$4333^{*}$ & $3433$ & $3343$ & $3334^{*}$ & $i$ & $-i$ & $i$ & $i$ &
$75.8/90^{*}$ & $76/90$ \\ \hline
\ET \EC}
\end{document}